\RequirePackage{lineno}
\documentclass[aps,prl,twocolumn,superscriptaddress,showpacs,preprintnumbers,amsmath,amssymb]{revtex4}
\usepackage{graphicx} % Include figure files
\usepackage{dcolumn}  % Align table columns on decimal point
\usepackage{bigstrut}
\graphicspath{{ps}}
\usepackage[format=plain,justification=RaggedRight,singlelinecheck=off]{caption}
\usepackage[format=plain,justification=justified,singlelinecheck=off]{subfig}

\usepackage{color}%\remove package when accepted

\begin{document}
%\linenumbers

%\vspace*{-3\baselineskip}
%\resizebox{!}{3cm}{\includegraphics{belle.eps}}

%\preprint{\vbox{ \hbox{   }
%			\hbox{Belle DRAFT {\it YY-NN}}
%                        \hbox{Intended for {\it PRD-RC}}
%                        \hbox{Author: C. Bele\~{n}o}
%                        \hbox{Committee: C. Schwanda(chair),}
%                        \hbox{A. Frey, Y. Watanabe }
%  		              % \hbox{hep-ex nnnn}
%}}

\title{\quad\\[1.0cm] Measurement of the Decays
  $\boldsymbol{B\to\eta\ell\nu_\ell}$ and
  $\boldsymbol{B\to\eta^\prime\ell\nu_\ell}$ in Fully Reconstructed
  Events at Belle}

%%%% >>>>> insert the authorlist here. BEFORE the abstract !!!!! <<<<<
%%%% >>>>> from the authorship confirmation web page
%%% Name the file author.tex and use \input{author} to insert into your latex file.

%%% Paper:    B -> eta(') l nu
\affiliation{Indian Institute of Technology Madras, Chennai 600036}
%%%\affiliation{Indiana University, Bloomington, Indiana 47408}
\affiliation{Institute of High Energy Physics, Chinese Academy of Sciences, Beijing 100049}
\affiliation{Institute of High Energy Physics, Vienna 1050}
%%%\affiliation{Institute for High Energy Physics, Protvino 142281}
%%%\affiliation{Institute of Mathematical Sciences, Chennai 600113}
\affiliation{INFN - Sezione di Torino, 10125 Torino}
\affiliation{Advanced Science Research Center, Japan Atomic Energy Agency, Naka 319-1195}
\affiliation{J. Stefan Institute, 1000 Ljubljana}
\affiliation{Kanagawa University, Yokohama 221-8686}
\affiliation{Institut f\"ur Experimentelle Kernphysik, Karlsruher Institut f\"ur Technologie, 76131 Karlsruhe}
%%%\affiliation{Kavli Institute for the Physics and Mathematics of the Universe (WPI), University of Tokyo, Kashiwa 277-8583}
\affiliation{Kennesaw State University, Kennesaw, Georgia 30144}
\affiliation{King Abdulaziz City for Science and Technology, Riyadh 11442}
\affiliation{Department of Physics, Faculty of Science, King Abdulaziz University, Jeddah 21589}
\affiliation{Korea Institute of Science and Technology Information, Daejeon 305-806}
\affiliation{Korea University, Seoul 136-713}
%%%\affiliation{Kyoto University, Kyoto 606-8502}
\affiliation{Kyungpook National University, Daegu 702-701}
\affiliation{\'Ecole Polytechnique F\'ed\'erale de Lausanne (EPFL), Lausanne 1015}
\affiliation{P.N. Lebedev Physical Institute of the Russian Academy of Sciences, Moscow 119991}
\affiliation{Faculty of Mathematics and Physics, University of Ljubljana, 1000 Ljubljana}
\affiliation{Ludwig Maximilians University, 80539 Munich}
\affiliation{Luther College, Decorah, Iowa 52101}
\affiliation{University of Malaya, 50603 Kuala Lumpur}
\affiliation{University of Maribor, 2000 Maribor}
\affiliation{Max-Planck-Institut f\"ur Physik, 80805 M\"unchen}
\affiliation{School of Physics, University of Melbourne, Victoria 3010}
%%%\affiliation{Middle East Technical University, 06531 Ankara}
%%% Journal:  Physical Review D Rapid Communications
%%% Contacts: C. Beleno (cesar-augusto.beleno-de-la-barrea@phys.uni-goettingen.de)
%%% Non-responding authors or those who said NO are commented out.
%%% ====================================================================
%%% Click the RELOAD button on your web browser to see the updated file.
%%% ====================================================================
%%% Use \input{author} to insert this material into your latex file.
%%%%% Force institutions to appear in alphabetical order when typeset.
\noaffiliation
%%%\affiliation{Aligarh Muslim University, Aligarh 202002}
\affiliation{University of the Basque Country UPV/EHU, 48080 Bilbao}
%%%\affiliation{Beihang University, Beijing 100191}
%%%\affiliation{University of Bonn, 53115 Bonn}
\affiliation{Budker Institute of Nuclear Physics SB RAS, Novosibirsk 630090}
\affiliation{Faculty of Mathematics and Physics, Charles University, 121 16 Prague}
%%%\affiliation{Chiba University, Chiba 263-8522}
\affiliation{Chonnam National University, Kwangju 660-701}
\affiliation{University of Cincinnati, Cincinnati, Ohio 45221}
\affiliation{Deutsches Elektronen--Synchrotron, 22607 Hamburg}
\affiliation{University of Florida, Gainesville, Florida 32611}
%%%\affiliation{Department of Physics, Fu Jen Catholic University, Taipei 24205}
%%%\affiliation{Justus-Liebig-Universit\"at Gie\ss{}en, 35392 Gie\ss{}en}
%%%\affiliation{Gifu University, Gifu 501-1193}
\affiliation{II. Physikalisches Institut, Georg-August-Universit\"at G\"ottingen, 37073 G\"ottingen}
\affiliation{SOKENDAI (The Graduate University for Advanced Studies), Hayama 240-0193}
\affiliation{Gyeongsang National University, Chinju 660-701}
\affiliation{Hanyang University, Seoul 133-791}
\affiliation{University of Hawaii, Honolulu, Hawaii 96822}
\affiliation{High Energy Accelerator Research Organization (KEK), Tsukuba 305-0801}
\affiliation{J-PARC Branch, KEK Theory Center, High Energy Accelerator Research Organization (KEK), Tsukuba 305-0801}
%%%\affiliation{Hiroshima Institute of Technology, Hiroshima 731-5193}
\affiliation{IKERBASQUE, Basque Foundation for Science, 48013 Bilbao}
%%%\affiliation{University of Illinois at Urbana-Champaign, Urbana, Illinois 61801}
%%%\affiliation{Indian Institute of Science Education and Research Mohali, SAS Nagar, 140306}
\affiliation{Indian Institute of Technology Bhubaneswar, Satya Nagar 751007}
\affiliation{Indian Institute of Technology Guwahati, Assam 781039}
\affiliation{University of Miyazaki, Miyazaki 889-2192}
\affiliation{Moscow Physical Engineering Institute, Moscow 115409}
\affiliation{Moscow Institute of Physics and Technology, Moscow Region 141700}
\affiliation{Graduate School of Science, Nagoya University, Nagoya 464-8602}
\affiliation{Kobayashi-Maskawa Institute, Nagoya University, Nagoya 464-8602}
%%%\affiliation{Nara University of Education, Nara 630-8528}
\affiliation{Nara Women's University, Nara 630-8506}
\affiliation{National Central University, Chung-li 32054}
\affiliation{National United University, Miao Li 36003}
\affiliation{Department of Physics, National Taiwan University, Taipei 10617}
%%%\affiliation{H. Niewodniczanski Institute of Nuclear Physics, Krakow 31-342}
\affiliation{Nippon Dental University, Niigata 951-8580}
\affiliation{Niigata University, Niigata 950-2181}
%%%\affiliation{University of Nova Gorica, 5000 Nova Gorica}
\affiliation{Novosibirsk State University, Novosibirsk 630090}
\affiliation{Osaka City University, Osaka 558-8585}
%%%\affiliation{Osaka University, Osaka 565-0871}
\affiliation{Pacific Northwest National Laboratory, Richland, Washington 99352}
%%%\affiliation{Panjab University, Chandigarh 160014}
%%%\affiliation{Peking University, Beijing 100871}
\affiliation{University of Pittsburgh, Pittsburgh, Pennsylvania 15260}
%%%\affiliation{Punjab Agricultural University, Ludhiana 141004}
%%%\affiliation{Research Center for Electron Photon Science, Tohoku University, Sendai 980-8578}
%%%\affiliation{Research Center for Nuclear Physics, Osaka University, Osaka 567-0047}
\affiliation{Theoretical Research Division, Nishina Center, RIKEN, Saitama 351-0198}
%%%\affiliation{RIKEN BNL Research Center, Upton, New York 11973}
%%%\affiliation{Saga University, Saga 840-8502}
\affiliation{University of Science and Technology of China, Hefei 230026}
%%%\affiliation{Seoul National University, Seoul 151-742}
%%%\affiliation{Shinshu University, Nagano 390-8621}
\affiliation{Showa Pharmaceutical University, Tokyo 194-8543}
\affiliation{Soongsil University, Seoul 156-743}
%%%\affiliation{University of South Carolina, Columbia, South Carolina 29208}
\affiliation{Stefan Meyer Institute for Subatomic Physics, Vienna 1090}
\affiliation{Sungkyunkwan University, Suwon 440-746}
\affiliation{School of Physics, University of Sydney, New South Wales 2006}
\affiliation{Department of Physics, Faculty of Science, University of Tabuk, Tabuk 71451}
\affiliation{Tata Institute of Fundamental Research, Mumbai 400005}
\affiliation{Excellence Cluster Universe, Technische Universit\"at M\"unchen, 85748 Garching}
%%%\affiliation{Department of Physics, Technische Universit\"at M\"unchen, 85748 Garching}
\affiliation{Toho University, Funabashi 274-8510}
%%%\affiliation{Tohoku Gakuin University, Tagajo 985-8537}
\affiliation{Department of Physics, Tohoku University, Sendai 980-8578}
\affiliation{Earthquake Research Institute, University of Tokyo, Tokyo 113-0032}
\affiliation{Department of Physics, University of Tokyo, Tokyo 113-0033}
\affiliation{Tokyo Institute of Technology, Tokyo 152-8550}
\affiliation{Tokyo Metropolitan University, Tokyo 192-0397}
%%%\affiliation{Tokyo University of Agriculture and Technology, Tokyo 184-8588}
\affiliation{University of Torino, 10124 Torino}
%%%\affiliation{Toyama National College of Maritime Technology, Toyama 933-0293}
%%%\affiliation{Utkal University, Bhubaneswar 751004}
\affiliation{Virginia Polytechnic Institute and State University, Blacksburg, Virginia 24061}
\affiliation{Wayne State University, Detroit, Michigan 48202}
\affiliation{Yamagata University, Yamagata 990-8560}
\affiliation{Yonsei University, Seoul 120-749}
\author{C.~Bele\~{n}o}\affiliation{II. Physikalisches Institut, Georg-August-Universit\"at G\"ottingen, 37073 G\"ottingen} % Goettingen
\author{J.~Dingfelder}\affiliation{University of Bonn, 53115 Bonn} % Bonn
\author{P.~Urquijo}\affiliation{School of Physics, University of Melbourne, Victoria 3010} % Melbourne
% \author{A.~Abdesselam}\affiliation{Department of Physics, Faculty of Science, University of Tabuk, Tabuk 71451} % Tabuk
% \author{I.~Adachi}\affiliation{High Energy Accelerator Research Organization (KEK), Tsukuba 305-0801}\affiliation{SOKENDAI (The Graduate University for Advanced Studies), Hayama 240-0193} % KEK
% \author{K.~Adamczyk}\affiliation{H. Niewodniczanski Institute of Nuclear Physics, Krakow 31-342} % Krakow
  \author{H.~Aihara}\affiliation{Department of Physics, University of Tokyo, Tokyo 113-0033} % Tokyo
  \author{S.~Al~Said}\affiliation{Department of Physics, Faculty of Science, University of Tabuk, Tabuk 71451}\affiliation{Department of Physics, Faculty of Science, King Abdulaziz University, Jeddah 21589} % Tabuk
% \author{K.~Arinstein}\affiliation{Budker Institute of Nuclear Physics SB RAS, Novosibirsk 630090}\affiliation{Novosibirsk State University, Novosibirsk 630090} % BINP
% \author{Y.~Arita}\affiliation{Graduate School of Science, Nagoya University, Nagoya 464-8602} % Nagoya
  \author{D.~M.~Asner}\affiliation{Pacific Northwest National Laboratory, Richland, Washington 99352} % PNNL
% \author{T.~Aso}\affiliation{Toyama National College of Maritime Technology, Toyama 933-0293} % Toyama
% \author{H.~Atmacan}\affiliation{University of South Carolina, Columbia, South Carolina 29208} % SouthCarolina
% \author{V.~Aulchenko}\affiliation{Budker Institute of Nuclear Physics SB RAS, Novosibirsk 630090}\affiliation{Novosibirsk State University, Novosibirsk 630090} % BINP
  \author{T.~Aushev}\affiliation{Moscow Institute of Physics and Technology, Moscow Region 141700} % MIPT
  \author{R.~Ayad}\affiliation{Department of Physics, Faculty of Science, University of Tabuk, Tabuk 71451} % Tabuk
% \author{T.~Aziz}\affiliation{Tata Institute of Fundamental Research, Mumbai 400005} % Tata
  \author{V.~Babu}\affiliation{Tata Institute of Fundamental Research, Mumbai 400005} % Tata
  \author{I.~Badhrees}\affiliation{Department of Physics, Faculty of Science, University of Tabuk, Tabuk 71451}\affiliation{King Abdulaziz City for Science and Technology, Riyadh 11442} % Tabuk
% \author{S.~Bahinipati}\affiliation{Indian Institute of Technology Bhubaneswar, Satya Nagar 751007} % IITB
% ...
% Ljubljana
% \author{N.~Zwahlen}\affiliation{\'Ecole Polytechnique F\'ed\'erale de Lausanne (EPFL), Lausanne 1015} % Lausanne
\collaboration{The Belle Collaboration}

%\author{C.~Bele\~no}\affiliation{University of G\"{o}ttingen} 
%\author{J.~Dingfelder}\affiliation{University of Bonn}
%\author{P.~Urquijo}\affiliation{University of Melbourne}
%\collaboration{The Belle Collaboration}
\noaffiliation
%% end author list

\begin{abstract}
We report branching fraction measurements of the decays
$B^+\to\eta\ell^+\nu_\ell$ and $B^+\to\eta^\prime\ell^+\nu_\ell$ based
on 711~fb$^{-1}$ of data collected near the $\Upsilon(4S)$ resonance
with the Belle experiment at the KEKB asymmetric-energy $e^+e^-$
collider. This data sample contains 772 million $B\bar
B$~events. One of the two $B$~mesons is fully reconstructed in a
hadronic decay mode. Among the remaining (``signal-$B$'') daughters, we search for the
$\eta$~meson in two decay channels,
$\eta\to\gamma\gamma$ and $\eta\to\pi^+\pi^-\pi^0$, and reconstruct
the $\eta^{\prime}$~meson in
$\eta^\prime\to\eta\pi^+\pi^-$ with subsequent decay of the $\eta$
into $\gamma\gamma$. Combining the two $\eta$ modes and using an extended maximum likelihood, the $B^+\to\eta\ell^+\nu_\ell$ branching fraction is measured to be 
$(4.2\pm 1.1 (\rm stat.)\pm
0.3 (\rm syst.))\times 10^{-5}$. For
$B^+\to\eta^\prime\ell^+\nu_\ell$, we observe no significant signal and
set an upper limit of $0.72\times 10^{-4}$ at 90\% confidence level.
\end{abstract}


\maketitle

%%%% >>>> keep the final version single-spaced
\tighten

{\renewcommand{\thefootnote}{\fnsymbol{footnote}}}
\setcounter{footnote}{0}
The magnitude of the Cabibbo-Kobayashi-Maskawa matrix element
$|V_{ub}|$~\cite{KM,cabibbo} can be determined by \textit{inclusive} measurements sensitive to
the entire $b\to u\ell\nu_\ell$~rate in a given region of phase
space, or by \textit{exclusive} measurements of specific $b\to u$~ decays such as
$B\to\pi\ell\nu_\ell$. As both experimental and
theoretical uncertainties differ in the two approaches,
consistency between the inclusive and exclusive determinations of $|V_{ub}|$ is a crucial cross-check of our
understanding of the CKM mechanism. At present, inclusive and exclusive measurements of $|V_{ub}|$ disagree
by about three standard deviations~\cite{HFAG}. Precise measurements
of $B\to\eta\ell\nu_\ell$ and $B\to\eta^\prime\ell\nu_\ell$ rates will improve the inclusive signal modelling, since the lack of knowledge on all exclusive $b\to u\ell\nu$ 
decays is one of the contributions to the systematic
uncertainty~\cite{Ball_Jones}. Also, a measurement of 
the ratio
$\mathcal{B}(B\to\eta\ell\nu_\ell)/\mathcal{B}(B\to\eta^\prime\ell\nu_\ell)$
 determines the $\eta-\eta^\prime$ mixing angle and the
${F_+^{B\to\eta^{(\prime)}}}$ form factor~\cite{eta-kim,eta-singlet}
by constraining the gluonic singlet contribution to this form factor
in the LCSR calculation~\cite{Ball_Jones}. In this paper, we report
measurements of the branching fractions
$\mathcal{B}(B^+\to\eta\ell^+\nu_\ell)$ and
$\mathcal{B}(B^+\to\eta^\prime\ell^+\nu_\ell)$~\cite{CC}, where
$\ell$ stands for either an electron or a muon. These are the first
measurements of these decays based on the Belle data sample. The
modes have been studied previously by CLEO~\cite{CLEOetapub,CLEOetapub2} and
BaBar~\cite{BABARetapub,BABARetapub2,BABARetapub3,BABARetapub4}.

The Belle detector is a large-solid-angle magnetic spectrometer that
consists of a silicon vertex detector (SVD), a 50-layer central drift
chamber (CDC), an array of aerogel threshold Cherenkov counters (ACC),
% <- \v{C}erenkov 2007.08
a barrel-like arrangement of time-of-flight scintillation counters
(TOF), and an electromagnetic calorimeter comprised of CsI(Tl)
crystals (ECL) located inside  a super-conducting solenoid coil that
provides a 1.5~T magnetic field.  An iron flux-return located outside
of the coil is instrumented to detect $K_L^0$ mesons and to identify
muons (KLM). The detector is described in detail
elsewhere~\cite{Belle}.

In this analysis, we use the entire Belle data sample of 711~fb$^{-1}$
collected at the KEKB asymmetric-energy $e^+e^-$ collider~\cite{KEKB} at the center-of-mass (c.m.) energy of the
$\Upsilon(4S)$~resonance. The sample contains $(772\pm 11)\times 10^6$
$e^+e^-\to\Upsilon(4S)\to B\bar B$ events. Two inner detector
configurations were used in the course of the experiment. A 2.0~cm
beampipe and a 3-layer silicon vertex detector were used for the first sample
of $152 \times 10^6$ $B\bar B$~pairs, while a 1.5~cm beampipe, a 4-layer
silicon detector, and a small-cell inner drift chamber were used to
record the remaining $620\times 10^6$ $B\bar B$ pairs~\cite{svd2}.  

Monte Carlo (MC) simulated samples are generated using the
EvtGen~\cite{EVTGEN} package and the response of the detector is
modeled using GEANT3~\cite{GEANT3}. MC samples equivalent to about
five times the integrated luminosity are produced for $\Upsilon(4S)\to
B\bar B$ events and $e^+e^-\to q\bar q$ continuum events, where $q$
stands for a $u$, $d$, $s$ or $c$ quark. Simulated samples
containing the decay $b\to u\ell\nu$ equivalent to 20 times the
integrated luminosity are used in this analysis. In these samples, the
decays $B^+\to\eta\ell^+\nu_\ell$ and
$B^+\to\eta^\prime\ell^+\nu_\ell$ have been generated according to the
ISWG2~\cite{ISGW2} calculation of the form factors.

After selecting hadronic events ($\Upsilon(4S)\to B\bar B$, $e^+e^-\to
q\bar q$) based on the charged track multiplicity and the total visible
energy~\cite{gas_beam}, we reconstruct one $B$~meson ($B_{\rm tag}$) of the
$B\bar B$ pair in a hadronic decay mode using the Belle full
reconstruction software~\cite{Feindt} based on the NeuroBayes neural-network package~\cite{Feindt2}. A total of 1104 exclusive decay channels to charm mesons and 71 neural networks were employed to reconstruct $B_{\rm tag}$ whose quality is characterized by the NeuroBayes classifier ($O_{NB}$), which ranges from 0 to 1. We require that $\ln O_{\rm NB}>-8$ to ensure good quality of $B_{\rm tag}$. $B_{\rm tag}$ is identified using the beam-constrained mass,
$M_\mathrm{bc}=\sqrt{{E_\mathrm{beam}^*}^2-|\vec p_{B_\mathrm{tag}}^*|^2}$,
and the energy difference, $\Delta
E=E^*_{B_\mathrm{tag}}-E^*_\mathrm{beam}$, where $E^*_\mathrm{beam}$
is the energy of the colliding beam particles in the c.m.\ frame and
$E^*_{B_\mathrm{tag}}$ and $\vec p^*_{B_\mathrm{tag}}$ are the
reconstructed energy and three-momentum of the $B_\mathrm{tag}$~candidate
in the same reference system~\cite{natural-units}. For well-reconstructed candidates, $\Delta E$ peaks at zero and $M_\mathrm{bc}$ peaks at the nominal
$B$ mass; we retain events that satisfy $-0.1\;{\rm GeV}<\Delta E<0.05\;{\rm GeV}$ and
$5.27\;{\rm GeV}<M_\mathrm{bc}<5.29\;{\rm GeV}$. Finally, we select only the charged $B_{\rm tag}$ candidates since the signal mode only involves charged $B$ mesons.

The other $B$ meson in the event, $B_{\rm sig}$, is reconstructed using all charged particles and neutral clusters not associated with the $B_{\rm tag}$ candidate. Low-momentum particles, which spiral inside the CDC and pass
close to the interaction point, can lead to multiple reconstruction of
the same particle. Duplicate tracks are identified as pairs of tracks with momenta transverse to the beam direction below $275$~MeV, with a momentum difference below 100~MeV, and with an opening angle either below 15$^\circ$ or above 165$^\circ$. Whenever such pair is found, we select the track passing closer to the interaction point.

Charged hadrons are identified using the ionization energy
loss $dE/dx$ in the CDC, the time-of-flight information provided by
the TOF, and the response of the ACC~\cite{Nakano:2002jw}. Pions used in this analysis are identified with an efficiency of 98\% and a kaon fake rate of 30\%.
Electron candidates are identified using the ratio of the energy detected in
the ECL to the track momentum, the ECL shower shape, the position matching
between the track and the ECL cluster, the energy loss in the CDC, and the
response of the ACC. Muons are identified based on their
penetration range and transverse scattering in the KLM detector. In
the momentum region relevant to this analysis, charged leptons are
identified with an efficiency of about 90\% and the probability to
misidentify a pion as an electron (muon) is 0.25\%
(1.4\%)~\cite{hanagaki,abashian}. We veto charged leptons
from photon conversion and $J/\psi$~decay if the lepton candidate, when combined with an oppositely charged particle, gives an invariant mass below 100~MeV or within $\pm 4.9$~MeV
around the nominal $J/\psi$~mass. Only events with a single charged
lepton candidate on the signal side are considered in this analysis.

Photons are reconstructed from clusters in the ECL not
matched to a track. Beam-related background is
removed by rejecting clusters with an energy below 50 MeV. Higher
thresholds of 100 MeV and 150 MeV are applied in the forward
($17^\circ<\theta<32^\circ$) and backward
($130^\circ<\theta<150^\circ$) regions, respectively, where
$\theta$ is the laboratory-frame polar angle with respect to the opposite of the positron beam
direction. Neutral pion candidates are reconstructed by combining two
photons, requiring their invariant mass to lie between 120 and
150~MeV. The c.m.\ momentum of the $\pi^0$~candidate must exceed
200~MeV.

Then, $\eta$~mesons are reconstructed in the decays $\eta\to\gamma\gamma$
and $\eta\to\pi^+\pi^-\pi^0$. Candidates are selected in the intervals
$0.506~\mathrm{GeV}<M_{\gamma\gamma}<0.584$~GeV and
$0.535~\mathrm{GeV}<M_{\pi^+\pi^-\pi^0}<0.560$~GeV, determined by
signal-to-background optimization on MC simulated events. We reconstruct
$\eta^\prime$ candidates in the $\eta^\prime\to\eta\pi^+\pi^-$ channel
with $\eta\to \gamma\gamma$ and require $0.926~\mathrm{GeV}<M_{\eta\pi^+\pi^-}<0.986$~GeV. The aforementioned mass requirements correspond to $3\sigma$ windows around the nominal mass of the mesons. The fraction of events with multiple meson candidates after the signal selection corresponds to $17.5\%$ for $\eta \to \gamma \gamma$, $7.4\%$ for $\eta\to\pi^{+}\pi^{-}\pi^{0}$ and $36\%$ for $\eta^{\prime}\to\eta(\gamma\gamma)\pi^{+}\pi^{-}$. If
more than one $\eta^{(\prime)}$~candidate is found on the signal side,
we select the one closer to the nominal
$\eta^{(\prime)}$~mass~\cite{PDG}. For modes involving charged pions,
we also use information on the signal vertex quality, and choose the candidate with the smallest $\chi^{2}_{\rm tot}$ defined as
$\chi^{2}_{\rm mass}+\chi^{2}_{\rm vertex}$. 

After selecting the single charged lepton and the $\eta^{(\prime)}$ candidate,
the remaining particles on the signal side are considered further to
reduce background. We require no remaining charged particles. The sum
of the energies of neutral clusters associated with neither 
$B_\mathrm{tag}$ nor $B_\mathrm{sig}$ must be below 0.5~GeV. To
reject charged leptons inconsistent with the signal decay, the charge of the lepton must be opposite to that of the
$B_{\rm tag}$ meson. Since the $\eta\to\gamma\gamma$ mode has a larger background than the $\eta\to\pi^{+}\pi^{-}\pi^{0}$ mode, we remove any events in the former channel that contain one or more neutral pions on the signal side. This $\pi^{0}$ veto is not applied to the $\eta^\prime\to\eta(\gamma\gamma)\pi^{+}\pi^{-}$ channel. 
\begin{figure*}[htpb]
%\begin{center}
%\hspace{-0.5cm}
\centering
%\null\hfill
\subfloat[$\eta\rightarrow\gamma\gamma$]{
\includegraphics[width=0.5\textwidth ]{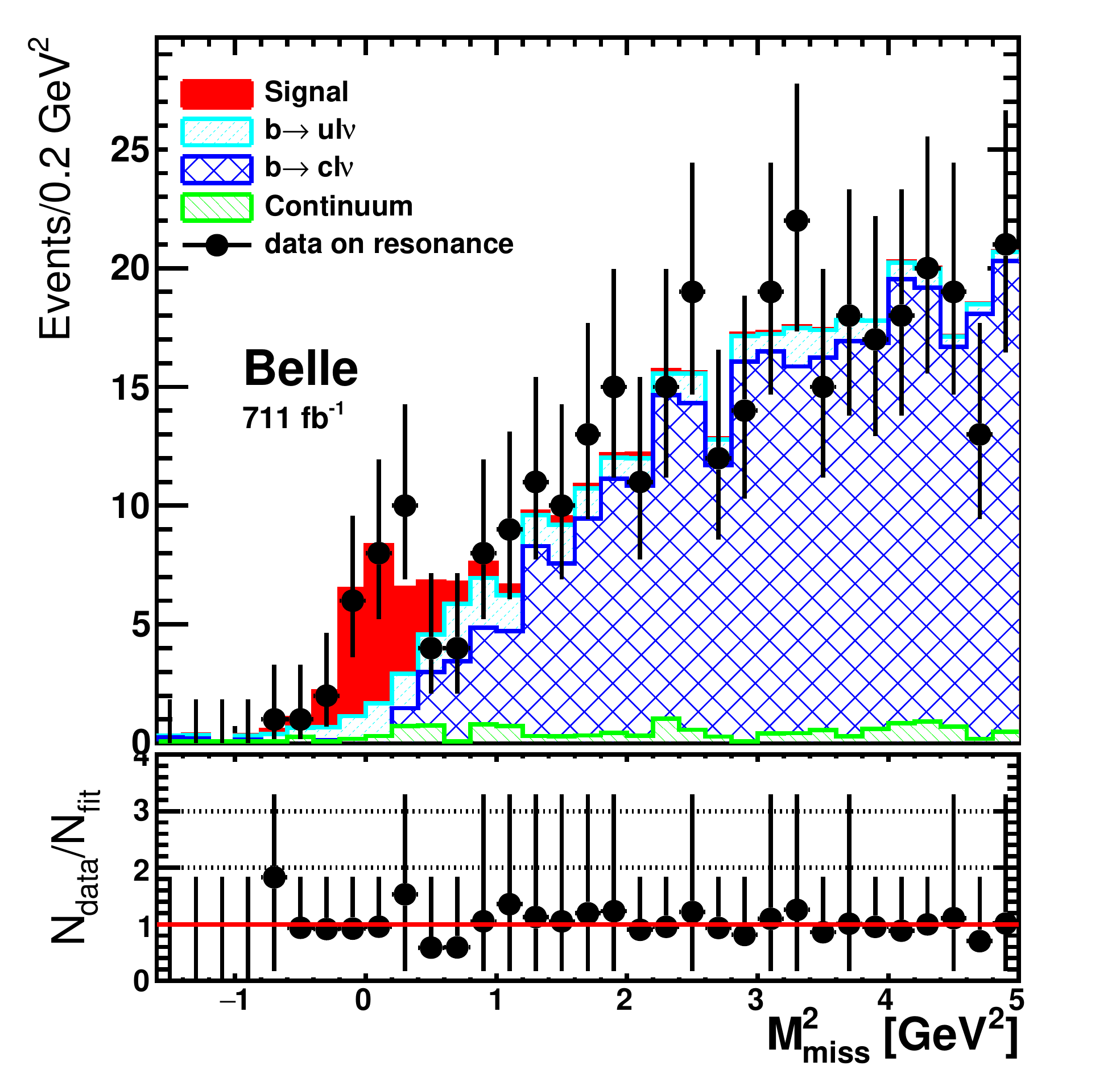}
}
%\hspace{-0.5cm}
\subfloat[$\eta\rightarrow \pi^{+}\pi^{-}\pi^{0}$]{
\includegraphics[width=0.5\textwidth ]{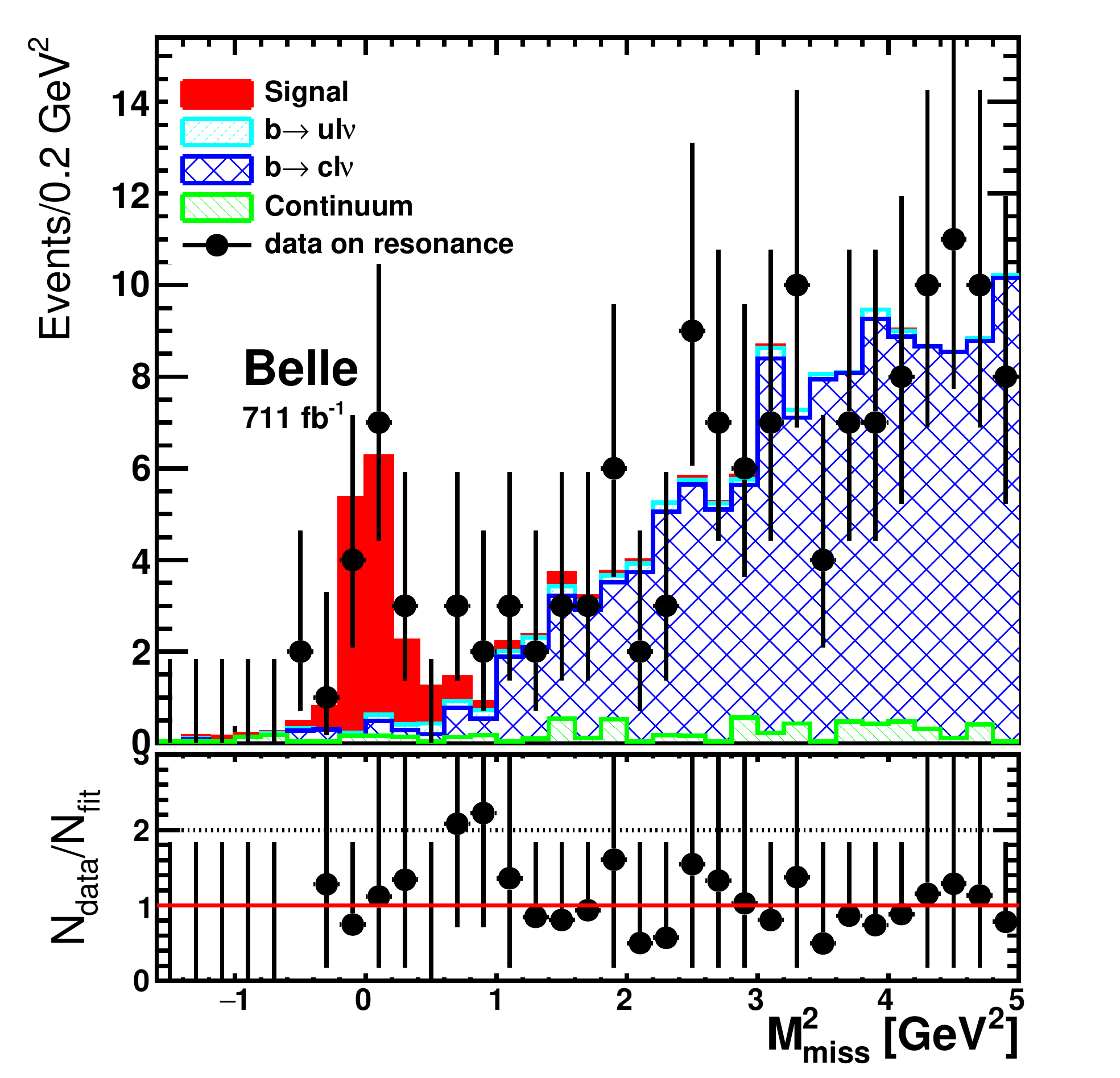}
}\\
%\vspace{-0.2cm}
%\hspace{-1cm}
\subfloat[Both $\eta$ modes combined]{
\includegraphics[width=0.5\textwidth ]{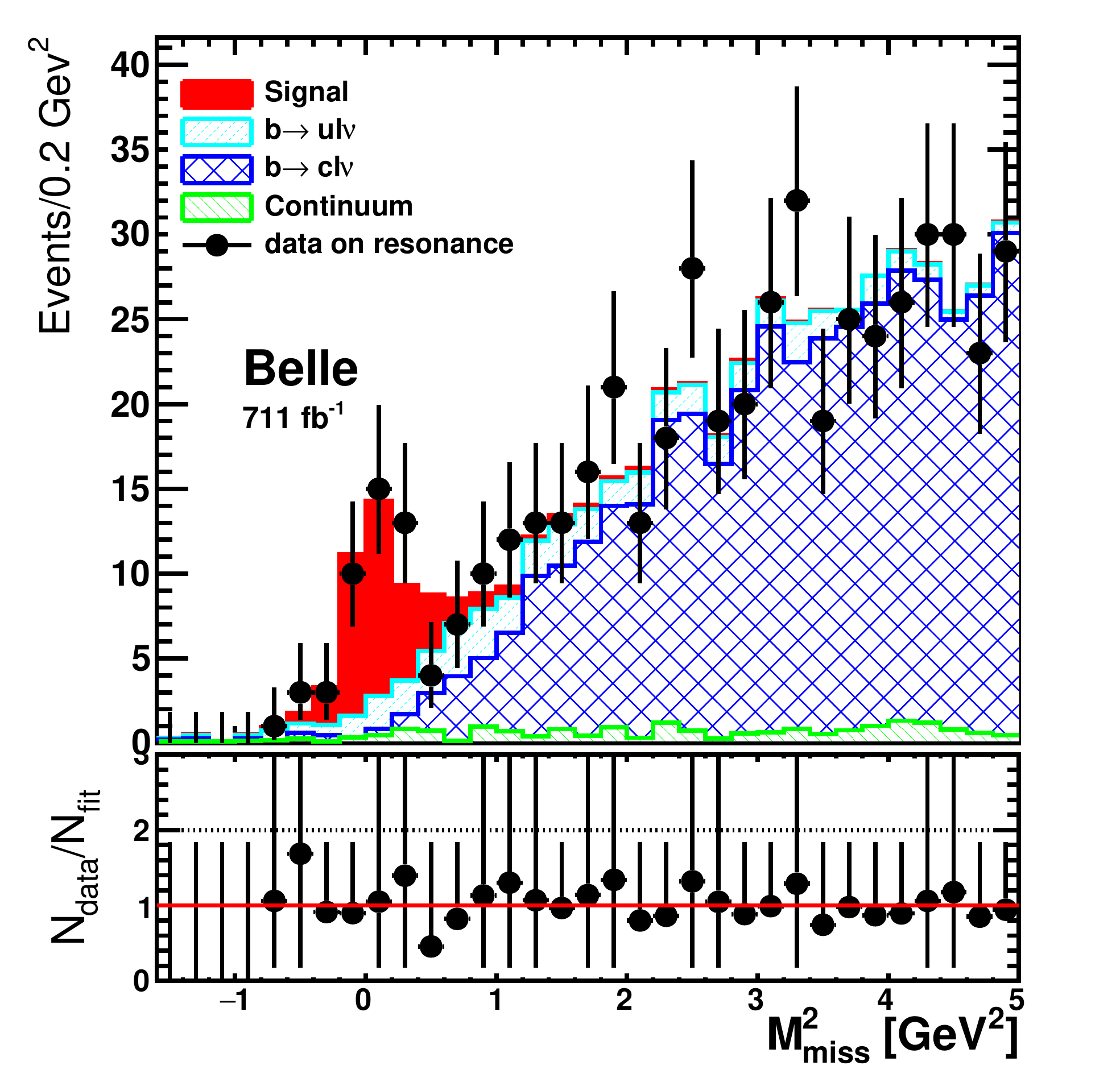}
}
\subfloat[$\eta^\prime \rightarrow \eta ( \gamma \gamma ) \pi^{+} \pi^{-}$]{
\includegraphics[width=0.5\textwidth]{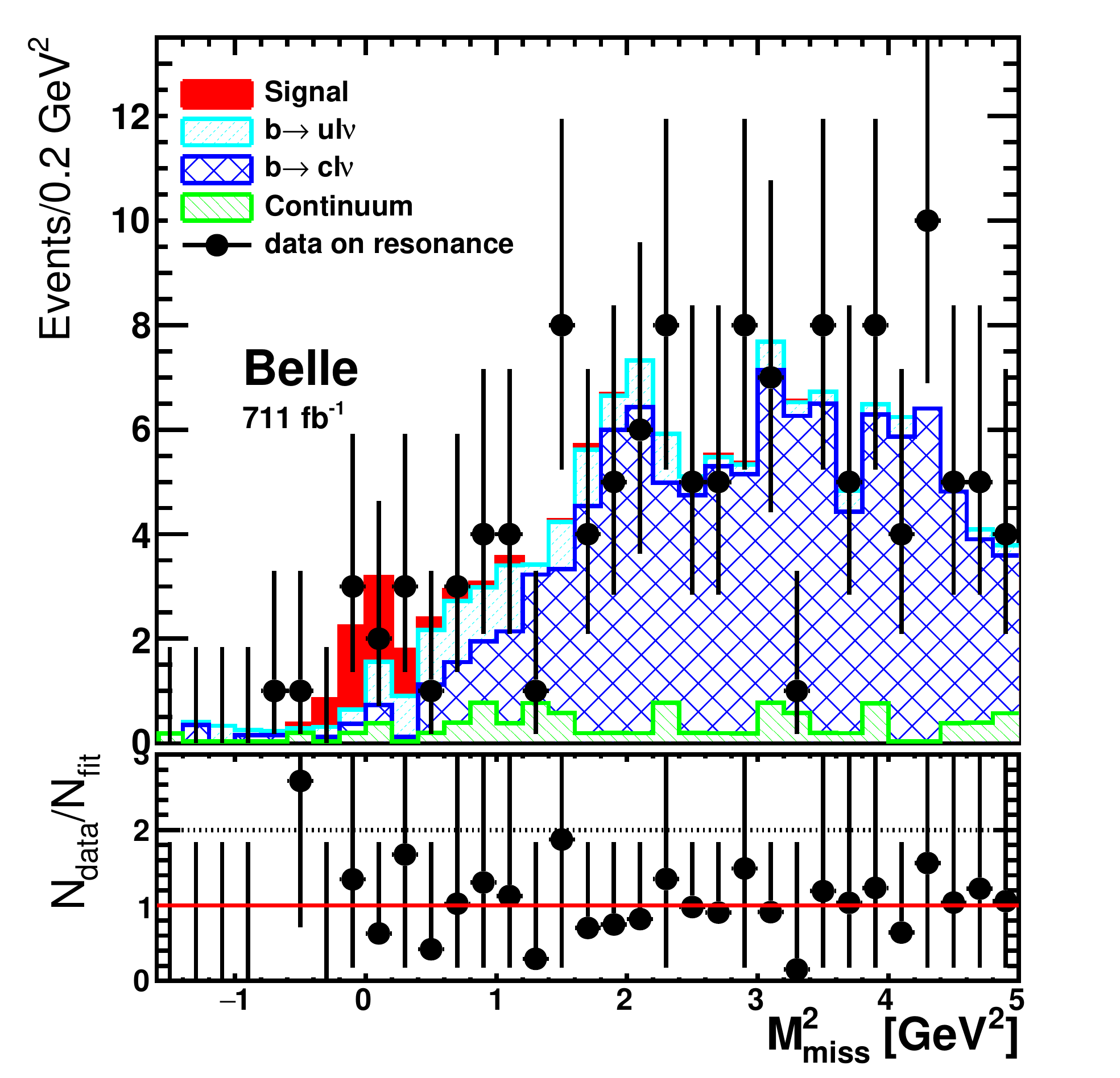}
}
%\hspace*{-1cm}
 % \hfill\null
\caption{Distribution of $M^2_\mathrm{miss}$ (points with error bars) for: (a)
  $\eta\to\gamma\gamma$, (b) $\eta\to\pi^+\pi^-\pi^0$, (c) both
  $\eta$~modes combined, and (d)
  $\eta^\prime\to\eta(\gamma\gamma)\pi^+\pi^-$. The fit results with the different components are shown as the colored histograms. The ratio of data to the sum of the fitted yields is shown below each plot.}
\label{fig:fitmmiss2}
%\end{center}
\end{figure*} 

\begin{table*}[htb]
%\captionsetup{justification=centerlast}
 \caption{Fit results in regions of $q^2=(p_\ell+p_{\nu_\ell})^2$ for
   the different modes. ``Raw yield'' denotes the number of events seen
   in the data; ``signal'', ``$b\to u\ell \nu_{\ell}$'', ``$b\to c\ell
   \nu_{\ell}$'' and ``continuum'' are the fitted yields. The continuum
   component is fixed and hence no fit errors are
   quoted. Only statistical uncertainties are shown.} \label{Tab:fit_results}
%\begin{center}
\small{
\hfill{}
\begin{tabular}
%{ @{\hspace{0.2cm}}l@{\hspace{0.2cm}}|  @{\hspace{0.2cm}}c@{\hspace{0.2cm}} @{\hspace{0.2cm}}c@{\hspace{0.2cm}} @{\hspace{0.2cm}}c@{\hspace{0.2cm}}|  @{\hspace{0.2cm}}c@{\hspace{0.2cm}}  
%@{\hspace{0.2cm}}c@{\hspace{0.2cm}}  @{\hspace{0.2cm}}c@{\hspace{0.2cm}}  @{\hspace{0.2cm}}c@{\hspace{0.2cm}}  @{\hspace{0.2cm}}c@{\hspace{0.2cm}}  @{\hspace{0.2cm}}c@{\hspace{0.2cm}}  |@{\hspace{0.2cm}}c@{\hspace{0.2cm}}   }
{l | ccc | ccc | ccc|c }
\hline\hline
Channel  & \multicolumn{9}{c|}{$\boldsymbol{B^+\to\eta\ell\nu_{\ell}}$} 
         & $\boldsymbol{B^+\to\eta^\prime\ell\nu_{\ell}}$\bigstrut  \\
\hline
 Mode    & \multicolumn{3}{c|}{$\boldsymbol{\eta\to\gamma\gamma}$} 
           &\multicolumn{3}{c|}{ $\boldsymbol{\eta\to\pi^+\pi^-\pi^0}$}
           &\multicolumn{3}{c|}{Both modes combined} &
 $\boldsymbol{\eta^\prime\to\eta(\gamma\gamma)\pi^+\pi^-}$\bigstrut \\

$q^{2}$ [GeV$\boldsymbol{^2}$]& All  & ${<12}$ & ${>12}$ 
                              & All  & ${<12}$ & ${>12}$ 
                              & All  & ${<12}$ & ${>12}$
                              & All   \bigstrut[b] \\
\hline
Raw yield 	                 &355   &261   &94 
                                                 &148   &98    &50   
                                                 &503   &359   &144     &129 \bigstrut[t]          \\
%%%%%%%%%%%%%%%%%%%%%%%%%%%%%%%%%%%%%%%%%%%%%%%%%%%%%%%%%%%%%%%%%%%%%%%%%%%%%%%%%%%%%%%%%%%%%%%%%%%%%%%%%%%%%%%%%%%%%
Signal                                          &23.6$\pm$8.7     &15.7$\pm$7.3     &9.0$\pm$5.3
                                                &16.0$\pm$5.3     &12.2$\pm$4.1     &4.0$\pm$2.5
                                                &38.8$\pm$10.1    &27.9$\pm$8.7     &12.9$\pm$6.1       &5.7$\pm$4.4\bigstrut  \\
%%%%%%%%%%%%%%%%%%%%%%%%%%%%%%%%%%%%%%%%%%%%%%%%%%%%%%%%%%%%%%%%%%%%%%%%%%%%%%%%%%%%%%%%%%%%%%%%%%%%%%%%%%%%%%%%%%%%%
%Signal corrected                                &24.3$\pm$9.2     &17.0$\pm$7.6     &10.0$\pm$6.1
%                                                &17.2$\pm$5.2     &13.1$\pm$4.6     &5.3$\pm$4.0
%                                                &40.7$\pm$10.6    &29.7$\pm$9.1     &14.7$\pm$7.0       &6.9$\pm$4.8\bigstrut  \\
%%%%%%%%%%%%%%%%%%%%%%%%%%%%%%%%%%%%%%%%%%%%%%%%%%%%%%%%%%%%%%%%%%%%%%%%%%%%%%%%%%%%%%%%%%%%%%%%%%%%%%%%%%%%%%%%%%%%%
$b\to u\ell \nu_{\ell}$  		         &32$\pm$25    &22$\pm$27   &10$\pm$13  
                                                 &4$\pm$21     &1$\pm$5     &4$\pm$8  
                                                 &46$\pm$29    &30$\pm$29   &14$\pm$18        &15$\pm$13\bigstrut  \\
%%%%%%%%%%%%%%%%%%%%%%%%%%%%%%%%%%%%%%%%%%%%%%%%%%%%%%%%%%%%%%%%%%%%%%%%%%%%%%%%%%%%%%%%%%%%%%%%%%%%%%%%%%%%%%%%%%%%%
$b\to c\ell\nu_{\ell}$ 		                 &287$\pm$27   &212$\pm$28  &73$\pm$13  
                                                 &122$\pm$17   &79$\pm$10   &41$\pm$10  
                                                 &399$\pm$31   &285$\pm$30  &114$\pm$18      &99$\pm$14\bigstrut  \\
%%%%%%%%%%%%%%%%%%%%%%%%%%%%%%%%%%%%%%%%%%%%%%%%%%%%%%%%%%%%%%%%%%%%%%%%%%%%%%%%%%%%%%%%%%%%%%%%%%%%%%%%%%%%%%%%%%%%%%
Continuum                                        &12.7         &10.4        &2.3  
                                                 &6.2          &5.2         &0.9 
                                                 &18           &15.7        &3.2               &$9.7$\bigstrut[b] \\
%%%%%%%%%%%%%%%%%%%%%%%%%%%%%%%%%%%%%%%%%%%%%%%%%%%%%%%%%%%%%%%%%%%%%%%%%%%%%%%%%%%%%%%%%%%%%%%%%%%%%%%%%%%%%%%%%%%%%%
\hline
$\epsilon$ $[10^{-3}]$                           &1.21         &1.28        &0.99  
                                                 &0.53         &0.57        &0.44 
                                                 &0.96         &1.02        &0.79               
                                                 &0.61\bigstrut[t] \\
%%%%%%%%%%%%%%%%%%%%%%%%%%%%%%%%%%%%%%%%%%%%%%%%%%%%%%%%%%%%%%%%%%%%%%%%%%%%%%%%%%%%%%%%%%%%%%%%%%%%%%%%%%%%%%%%%%%%%%
\hline
$\chi^2/$ndf		                         &$12.0/29$   &$11.5/29$   &$35.2/29$
                                                 &$18.8/29$   &$30.5/29$   &$19.4/29$  
                                                 &$18.0/29$   &$20.1/29$   &$35.5/29$        &$24.4/29$\bigstrut[t]  \\
%%%%%%%%%%%%%%%%%%%%%%%%%%%%%%%%%%%%%%%%%%%%%%%%%%%%%%%%%%%%%%%%%%%%%%%%%%%%%%%%%%%%%%%%%%%%%%%%%%%%%%%%%%%%%%%%%%%%%
Probability[\%]		                         &$99.8$      &$99.8$      &$19.0$ 
                                                 &$92.5$      &$39.1$      &$91.1$
                                                 &$94.4$      &$88.8$      &$18.9$           &$70.9$\bigstrut[b]  \\
\hline\hline
\end{tabular}}
\hfill{}
\end{table*}

The $B\to\eta^{(\prime)}\ell\nu_\ell$ yield is extracted from the
distribution of the missing mass squared, defined as
$M^2_\mathrm{miss}=(p_{B_\mathrm{tag}}-p_{\eta^{(\prime)}}-p_\ell)^2$,
where $p_{B_\mathrm{tag}}$, $p_{\eta^{(\prime)}}$ and $p_\ell$ are the
four-momenta of the $B_\mathrm{tag}$, $\eta^{(\prime)}$, and charged
lepton candidates, respectively. For well-reconstructed signal decays,
we expect $M^2_\mathrm{miss}$ to peak at zero, as the only
remaining particle in the event is the neutrino. We determine the
yields of the signal, $b\to u\ell\nu_\ell$, $b\to c\ell\nu_\ell$ and continuum backgrounds from an
extended binned maximum likelihood fit to the $M^2_\mathrm{miss}$
distribution between $-1.6$ and 5.0~GeV$^2$ (with a bin width of 0.2
GeV$^2$). The shapes of the fit components are taken from MC simulation and the
fitting algorithm accounts for statistical fluctuations in both the
real data and the MC simulated samples~\cite{barlow_beeston}. As continuum
is a small component, we fix it to the MC
expected yield. The contributions from secondary and fake leptons are negligible and thus not taken into account as additional fit components. For $B^+\to\eta\ell\nu_{\ell}$, the fit incorporates both $\eta$~modes. As a cross-check, we also determine the
fit results for the individual $\eta$~modes. In
addition, we include also fit results for the regions of
$q^2=(p_\ell+p_{\nu_\ell})^2$ below and above 12~GeV$^2$. These fit
results are quoted in Table~\ref{Tab:fit_results} and shown in
Fig.~\ref{fig:fitmmiss2}. We carried out 10000 toy MC to validate the fit procedure. The distributions of signal and background in each ensemble are generated according to their measured values in data, and then the fit procedure is executed. The statistical uncertainties estimated by the nominal fit are consistent with the size of the uncertainties evaluated by the toy MC technique. However, given that in some channels the pull distribution exhibits a non-Gaussian shape, we do not apply a correction to the central value of the signal yields. Instead, we assign a systematic uncertainty associated with the fit procedure with values between 2\% and 10\% depending on the reconstructed channel. 

%We created 1000 toy Monte Carlo samples based on the missing mass squared histograms, by fluctuating the distribution for each input Monte Carlo sample bin by bin using Poisson statistics. The Monte Carlo sample components are added to make up the total distribution that corresponds to the toy data, which is then fitted to study the pull distribution. A stable, Gaussian distributed fit result is found for the different modes with a mean of the pull close to zero and a variance  consistent with one. This validation procedure ensures that the technique is unbiased and stable, and also that the errors are estimated correctly.

The signal branching fractions are calculated as
\begin{equation}
  \mathcal{B}(B^+\to\eta^{(\prime)}\ell^+\nu_{\ell})=\frac{1}{2}\frac{N_\mathrm{signal}}{N_{B\bar
      B}\mathcal{B}(\eta^{(\prime)})\epsilon}~,
\end{equation}
where $N_\mathrm{signal}$ is the fitted signal yield from
Table~\ref{Tab:fit_results}, $N_{B\bar B}$ is the number of $B\bar B$
pairs in the Belle data, $\mathcal{B}(\eta^{(\prime)})$ is the
world average value of the $\eta^{(\prime)}$ sub-decay branching
fraction~\cite{PDG,combined} and $\epsilon$ is the signal efficiency
including $B_{\rm tag}$ reconstruction, calibrated as described in
Ref.~\cite{sibidanov}. The factor of 2 in the denominator indicates an average over lepton flavor. The combined and separate $B^+\to\eta\ell^+\nu_\ell$ branching
fractions  are quoted in
Table~\ref{tab:Branching_fraction}. Our result for the
$B^+\to\eta\ell^+\nu_\ell$ branching fraction is $(4.2\pm 1.1 (\rm stat)\pm
0.3 (\rm syst.))\times 10^{-5}$. The significance of the observed
signal~\cite{FeldmanCousins,significance} is calculated as
$S=\sqrt{-2\Delta\ln({\cal L})}$ with $\Delta\ln({\cal L})=\ln({\cal
  L}_{\rm B})-\ln({\cal L}_{\rm S+B})$, where $\ln({\cal L}_{\rm
  S+B})$ is the maximized log-likelihood assuming a signal plus
background hypothesis and $\ln({\cal L}_{\rm B})$ is the
maximized log-likelihood with background only. Systematic uncertainties are
included by convolving ${\cal L}$ with a Gaussian function of width
corresponding to the systematic uncertainty in the number of signal
events. The signal significance in the combined $\eta$~mode sample is
found to be $S=3.7$, including systematic uncertainties related to the signal yield.

\begin{table*}[htb]
%\captionsetup{justification=centerlast}
\caption{Branching fraction of the decay $B^+\to\eta\ell^+\nu_\ell$
  (in units of $10^{-5}$) calculated for the different samples and regions
  of $q^2$. The first error is statistical and the second is
  systematic. The main result is the lower right value (Combined, All $q^{2}$). The values in the ``Sum'' row provide a cross-check.} \label{tab:Branching_fraction}
\begin{center}
\begin{tabular}{l c c c }
\hline\hline
                       & $\eta\to\gamma\gamma$ & $\eta\to\pi^+\pi^-\pi^0$& Combined\bigstrut[t]\\
\hline
$q^2<12$~GeV$^2$             &$2.0\pm0.9\pm 0.2$ 
                             &$6.0\pm2.0\pm 0.6$ 
                             &$2.8\pm0.9\pm 0.2$ \bigstrut[b]\\
$q^2>12$~GeV$^2$            &$1.5\pm0.9\pm 0.1$
                             &$2.6\pm1.6\pm 0.3$ 
                             &$1.7\pm0.8\pm 0.1$ \bigstrut[t]\\

Sum                              &$3.5 \pm 1.3 \pm 0.3 $
                                 &$8.6\pm2.6 \pm 0.7  $ 
                                 &$4.5\pm1.2 \pm 0.3$  \bigstrut[b]\\
\hline
All $q^2$                        &$3.4 \pm 1.2 \pm 0.3 $
                                 &$8.5\pm2.8 ^{+0.7}_{-0.8}  $ 
                                 &\hbox{\boldmath$4.2\pm1.1 \pm 0.3$}  \bigstrut[b]\\
\hline\hline
\end{tabular}
\end{center}
\end{table*}

For $B^+\to\eta^\prime\ell^+\nu_\ell$, we calculate a branching fraction
of $(3.6\pm 2.7 (\rm stat.)^{+0.3}_{-0.4}(\rm syst.))\times 10^{-5}$ and a
significance (including systematics) of $S=1.6$. Given the low
value of $S$, we convert this result into an upper limit on
$\mathcal{B}(B^+\to\eta^\prime\ell^+\nu_\ell)$. Using the frequentist
calculator from the RooStats package~\cite{RooStats}, we obtain a 90\% confidence level upper limit of 11.6 events on the
$B^+\to\eta^\prime\ell^+\nu_\ell$ signal yield or
$0.72\times 10^{-4}$ on the branching fraction. For the $B^{+}\to\eta\ell^+\nu_\ell$ channel this upper limit is of 51.2 events, corresponding an upper bound on the branching ratio of $0.55\times 10^{-4}$. 

We also determine the ratio ${\cal B}(B^{+}\to\eta^\prime\ell^+\nu_\ell)/ {\cal B}(B^{+}\to\eta\ell^+\nu_\ell)$ to be $0.86\pm 0.68 (\rm stat.)\pm 0.09(\rm syst.)$, which is important to constraint the gluonic singlet contribution~\cite{Ball_Jones}. A 90\% confidence level upper limit to the latter quantity is calculated to be ${\cal B}(B^{+}\to\eta^\prime\ell^+\nu_\ell)/ {\cal B}(B^{+}\to\eta\ell^+\nu_\ell)<1.31$.

We compute the CKM matrix element $|V_{ub}|$ from our measurement of ${\cal B} (B^+\to\eta\ell^+\nu_\ell)$ in the region $q^{2}<12$ GeV$^{2}$ using the light-cone sum rule (LCSR) calculation of the form factor $f_{+}(q^{2})$ in Ref.~\cite{Ball_Jones}. For that purpose we use the relation:

\begin{equation}
|V_{ub}|=\sqrt{\frac{C_{v}\Delta{\cal B}}{\tau_{B}\Delta\zeta}},
\end{equation}

where $C_{v}=2$ for $B^{+}$ decays, $\Delta{\cal B}$ is the measured partial branching ratio for $q^{2}<12$ GeV$^{2}$, $\tau_{B}=1.638(4)$ ps~\cite{PDG} is the lifetime of the $B^{+}$ meson and $\Delta\zeta$ is the decay rate provided by theory~\cite{Ball_Jones}. We determine $\Delta\zeta$ to be $(2.65^{+0.43}_{-0.47})\times 10^{12}$ s$^{-1}$ and consequently $|V_{ub}|=(3.59\pm 0.58 (\rm stat.)\pm 0.13(\rm syst.) ^{+0.29}_{-0.32}(\rm theo.))\times 10^{-3}$, which is in agreement with previous exclusive measurements~\cite{HFAG}.

The systematic uncertainties considered for the branching fractions
are summarized in  Table~\ref{tab:systerrors} and fall into two groups:
those related to detector perfomance and those in the
signal and background modeling. Uncertainties related to detector performance
are derived from dedicated studies of control samples within the Belle experiment to measure
the tracking efficiency of charged particles, the photon and neutral-pion reconstruction efficiency, and the charged-lepton and pion-identification efficiency. Systematic uncertainties related to the
signal and background model are estimated by varying the respective
parameter in the simulation within its uncertainty or by reweighting
MC samples. The deviation of the result from the nominal fit is taken
as the uncertainty.

Uncertainties in the signal form factors are estimated by comparing
the Ball-Zwicky model~\cite{Ball_Zwicky} to the ISGW2~model~\cite{ISGW2}. The
form factor parameters of the former are taken from
Ref.~\cite{Ball_Jones}. The HQET-based form factors of the decays
$B\to D^{(*)}\ell\nu_{\ell}$ in the MC simulation are adjusted to the
recent world average values~\cite{HFAG}. The branching
fractions of $B\to (D^{(*)},\pi,\rho,\omega)\ell\nu_\ell$ have been
corrected~\cite{PDG}. The hadronic branching fractions on the tag side
are adjusted by the $B_\mathrm{tag}$~calibration and its uncertainty is
taken from Ref.~\cite{sibidanov}. We vary the branching fractions
of the $b\to u\ell\nu_{\ell}$ and $b\to
c\ell\nu_{\ell}$ decay modes within $\pm 1$ standard deviation of their world
average values. We consider the form-factor uncertainties in the
decays $B\to D^*\ell\nu_\ell$, $B\to D^0\ell\nu_\ell$,
$B\to\pi\ell\nu_\ell$, and $B\to\omega\ell\nu_\ell$, and uncertainties
in the shape-function parameters of the inclusive $b\to u\ell\nu_\ell$
model. We further assign an uncertainty due to the branching fraction
uncertainty in the $\eta^{(\prime)}$ sub-decay modes. The systematic
error components in which a weight factor is applied include uncertainties
due to secondary and fake leptons and the continuum. The contribution
of the secondary leptons is adjusted to the measured $b\to c\to\ell$
branching fraction. The contribution of events in which a lepton has
been misidentified as a hadron is corrected using the fake rate
measured in a kinematically selected $D^{*+}\to
D^0(K^-\pi^+)\pi^+$ sample. Since the expected number of continuum events is small after signal selection, a comparison with off-resonance data is not carried out. Instead, we rely on MC simulation to estimate the systematic uncertainty associated with continuum normalization by varying the number of events by 20\% and examining the effect on the fit. The deviation from the nominal fit is taken as the uncertainty. The uncertainty on the number of produced $B$-meson pairs is 1.4\%. 

%The off--resonance data was not employed in this analysis because of the small data size that lead to high statistical uncertainty. Finally, the determination of tagged $B$ mesons was performed using a control sample of $B\to D^{*0}\ell\nu$ decays. A three component fit to the $M_{\rm bc}$ distribution is carried out taking into account the shapes of the Monte Carlo samples. The fit components are: continuum, combinatorial $B$ background and well reconstructed $B$ signal. Here combinatorial $B$ background refers to any reconstructed $B$ for which a track has been lost or whose flavour or charge have been misidentified. Reconstructed $B$ decays with a lost photons and tracks with the incorrect particle ID hypothesis are considered signal. The number of well reconstructed $B^+$ decays is found to be $(1.77\pm 0.01\pm 0.09)\times 10^6$ with an associated systematic uncertainty of $4.5\%$. 

%$^{+}_{-}$
\begin{table*}[htpb]
\caption{Relative systematic uncertainties in the signal yield in per cent for the fits to
  the two $\eta$-mode samples and in the different $q^2$
  regions.} \label{tab:systerrors}
\begin{center}{\small
\hspace{-1.7cm}
\begin{tabular}{l |ccc |ccc| ccc |c}
\hline   
Mode	   &\multicolumn{3}{c|}{$\boldsymbol{\eta\to\gamma\gamma}$ }
                           & \multicolumn{3}{c|}{$\boldsymbol{\eta\to\pi^+\pi^-\pi^0}$}
                           &\multicolumn{3}{c|}{\bf{Both $\eta$ modes}}
                           &$\boldsymbol{\eta^\prime\to\eta(\gamma\gamma)\pi^+\pi^-}$\bigstrut[t] \\
$q^{2}$ [GeV$^2$]	   & All  & $<12$ & $>12$
                           & All  & $<12$ & $>12$
                           & All  & $<12$ & $>12$
                           & All \bigstrut[b] \\
\hline
Track finding 		     &$\pm0.35$&$\pm0.35$&$\pm0.35$ 		
                             &$\pm1.05$&$\pm1.05$&$\pm1.05$   	
                             &$\pm0.5$&$\pm0.5$&$\pm0.5$                               
                             &$\pm1.05$\bigstrut \\
Photon finding 		     &$\pm4.0$&$\pm4.0$&$\pm4.0$ 		
                             &$\pm0.0$&$\pm0.0$&$\pm0.0$		        
                             &$\pm3.1$&$\pm3.1$&$\pm3.1$                                     
                             &$\pm4.0$\bigstrut \\
$\pi^0$ reconstruction                        &$\pm0.0$&$\pm0.0$&$\pm0.0$		
                                              &$\pm2.5$&$\pm2.5$&$\pm2.5$		
                                              &$\pm0.5$&$\pm0.5$&$\pm0.5$                               
                                              &$\pm0.0$\bigstrut \\
$\pi^0$ veto                                  &$\pm2.5$&$\pm2.5$&$\pm2.5$		
                                              &$\pm0.0$&$\pm0.0$&$\pm0.0$		
                                              &$\pm2.0$&$\pm2.0$&$\pm2.0$                               
                                              &$\pm0.0$\bigstrut \\
Pion ID			     &$\pm0.0$&$\pm0.0$&$\pm0.0$		
                             &$\pm1.0$&$\pm1.0$&$\pm1.0$ 			
                             &$\pm0.20$&$\pm0.20$&$\pm0.20$                            
                             &$\pm1.0$\bigstrut \\
Lepton ID 		     &$\pm2.0$&$\pm2.0$&$\pm2.0$		
                             &$\pm2.0$&$\pm2.0$&$\pm2.0$ 	        
                             &$\pm2.0$&$\pm2.0$&$\pm2.0$                               
                             &$\pm2.0$\bigstrut \\
Lepton fake rate 	     &$\pm0.36$&$^{+0.19}_{-0.13}$&$\pm0.11$		
                             &$^{+0.46}_{-0.50}$&$^{+0.42}_{-0.47}$&$^{+0.18}_{-0.16}$ 	        
                             &$^{+0.47}_{-0.44}$&$\pm0.51$&$^{+0.02}_{-0.07}$                               
                             &$^{+1.6}_{-1.8}$\bigstrut[b] \\
\hline
Signal model              &$\pm0.83$&$\pm0.75$&$\pm1.0$		
                             &$\pm0.50$&$\pm0.70$&$\pm0.46$		
                             &$\pm0.88$&$\pm0.71$&$\pm2.0$                               
                             &$\pm0.28$\bigstrut \\
$b\to u\ell\nu_\ell$ form factors &$\pm1.1$&$\pm0.49$&$\pm0.72$ 		
                                    &$^{+1.8}_{-2.6}$&$^{+0.14}_{-0.16}$&$^{+0.82}_{-1.4}$ 	        
                                    &$^{+0.31}_{-0.43}$&$^{+0.73}_{-1.1}$&$^{+0.77}_{-0.70}$                               
                                    &$^{+0.92}_{-0.56}$\bigstrut[t] \\
$b\to u\ell\nu_\ell$ branching fractions &$^{+0.26}_{-0.20}$&$\pm1.0$&$^{+1.4}_{-1.3}$ 		
                                    &$^{+0.04}_{-0.05}$&$\pm0.05$&$^{+0.85}_{-0.95}$	
                                    &$^{+0.50}_{-0.45}$&$^{+1.5}_{-1.8}$&$^{+0.86}_{-1.2}$    
                                    &$^{+1.9}_{-2.4}$\bigstrut \\
$b\to c\ell\nu_\ell$ form factors    &$^{+1.0}_{-0.15}$&$^{+2.3}_{-0.60}$&$\pm0.0$		
                                    &$^{+0.21}_{-0.06}$&$^{+0.70}_{-0.22}$&$\pm0.0$		
                                    &$^{+1.1}_{-0.10}$&$^{+1.3}_{-0.24}$&$\pm0.0$                                 
                                    &$^{+0.18}_{-0.23}$\bigstrut \\
$b\to c\ell\nu_\ell$ branching fractions &$\pm0.14$&$\pm0.80$&$\pm0.29$  		
                                    &$\pm0.28$&$^{+0.43}_{-0.45}$&$^{+0.18}_{-0.28}$ 	
                                    &$\pm0.13$&$\pm0.64$&$^{+0.21}_{-0.27}$   
                                    &$\pm0.62$\bigstrut \\
Secondary leptons	     &$^{+0.00}_{-0.06}$&$\pm 0.12$&$^{+0.01}_{-0.03}$  		
                             &$^{+0.07}_{-0.04}$&$^{+0.15}_{-0.13}$&$^{+0.02}_{-0.12}$		
                             &$^{+0.03}_{-0.01}$&$\pm 0.08$&$^{+0.06}_{-0.04}$                              
                             &$^{+0.01}_{-0.00}$\bigstrut[t] \\
${\cal B}(\eta^{(\prime)})$~\cite{combined} &$\pm0.50$&$\pm0.50$&$\pm0.50$		
                                &$\pm1.2$&$\pm1.2$&$\pm1.2$		
                                &$\pm0.50$&$\pm0.50$&$\pm0.50$                               
                                &$\pm1.7$\bigstrut \\
Hadronic tag 		     & $\pm4.2$&$\pm4.2$&$\pm4.2$
                    	     &$\pm4.2$&$\pm4.2$&$\pm4.2$ 		
                             &$\pm4.2$&$\pm4.2$&$\pm4.2$                               
                             &$\pm4.2$\bigstrut[b] \\
N$(B\bar{B})$ 		     & $\pm1.4$&$\pm1.4$&$\pm1.4$
                    	     &$\pm1.4$&$\pm1.4$&$\pm1.4$ 		
                             &$\pm1.4$&$\pm1.4$&$\pm1.4$                               
                             &$\pm1.4$\bigstrut[b] \\
Continuum 		     &$^{+0.77}_{-0.80}$&$^{+0.98}_{-0.96}$&$^{+0.24}_{-0.30}$
                	     &$^{+0.66}_{-0.64}$&$^{+1.1}_{-1.2}$&$^{+0.71}_{-0.62}$
              	             &$\pm 0.47$&$\pm 0.83$&$^{+1.2}_{-1.3}$                               
                             &$\pm 3.9$\bigstrut \\
Fit procedure 		     &$\pm2.9$&$\pm9.8$&$\pm2.0$
                    	     &$\pm6.3$&$\pm8.7$&$\pm9.6$ 		
                             &$\pm2.2$&$\pm5.6$&$\pm3.2$                               
                             &$\pm5.2$\bigstrut[b] \\
\hline
Total                        &$\pm7.6$        &$^{+12.3}_{-12.1}$  &$\pm7.3$	
                             &$^{+8.8}_{-9.0}$&$\pm10.6$           &$^{+11.3}_{-11.4}$ 		
                             &$\pm6.7$        &$\pm8.7$            &$^{+7.4}_{-7.5}$
                             &$^{+9.7}_{-9.8}$\bigstrut \\
\hline
\end{tabular}

\hspace*{-1cm}}
\end{center}
\end{table*}

In summary, we have measured the branching fraction of the decay
$B^+\to\eta\ell^+\nu_\ell$ to be $(4.2\pm 1.1\pm
0.3)\times 10^{-5}$, where the first error is statistical and
the second systematic. For the branching fraction of
$B^+\to\eta^{\prime}\ell^+\nu_\ell$, we determine a 90\% confidence level
upper limit of $0.72\times 10^{-4}$. The measurements are compatible
with previous analyses performed by CLEO and BaBar~\cite{CLEOetapub,CLEOetapub2,BABARetapub,BABARetapub2,BABARetapub3,BABARetapub4}. Our
measurement is limited by the size of the Belle data
sample. Significant improvements can thus be expected from the Belle
II/SuperKEKB super flavor factory.
 
% This measurement of $Beauty$  
% is based on a data sample that
% contains $XXX \times 10^6 B\overline{B}$ pairs, 
% collected  with the Belle detector at the KEKB asymmetric-energy
% $e^+e^-$ (3.5 on 8~GeV) collider~\cite{KEKB}
% operating at the $\Upsilon(4S)$ resonance.

% {\it SVD1:}
% {The Belle detector is a large-solid-angle magnetic
% {spectrometer that
% {consists of a three-layer silicon vertex detector (SVD),
% {a 50-layer central drift chamber (CDC), an array of
% {aerogel threshold Cherenkov counters (ACC), % <- \v{C}erenkov 2007.08
% {a barrel-like arrangement of time-of-flight
% {scintillation counters (TOF), and an electromagnetic calorimeter
% {comprised of CsI(Tl) crystals (ECL) located inside 
% {a super-conducting solenoid coil that provides a 1.5~T
% {magnetic field.  An iron flux-return located outside of
% {the coil is instrumented to detect $K_L^0$ mesons and to identify
% {muons (KLM).  The detector
% {is described in detail elsewhere~\cite{Belle}.

% {\it SVD2+SVD1:}

%The procedure for flavor tagging is described in Ref.~\cite{TaggingNIM}
%while the improved continuum suppression algorithm is discussed in Ref.~\cite{KSFW}.

%***** Acknowledgments *****

We thank the KEKB group for the excellent operation of the
accelerator; the KEK cryogenics group for the efficient
operation of the solenoid; and the KEK computer group,
the National Institute of Informatics, and the 
PNNL/EMSL computing group for valuable computing
and SINET5 network support.  We acknowledge support from
the Ministry of Education, Culture, Sports, Science, and
Technology (MEXT) of Japan, the Japan Society for the 
Promotion of Science (JSPS), and the Tau-Lepton Physics 
Research Center of Nagoya University; 
the Australian Research Council;
Austrian Science Fund under Grant No.~P 26794-N20;
the National Natural Science Foundation of China under Contracts 
No.~10575109, No.~10775142, No.~10875115, No.~11175187, No.~11475187, 
No.~11521505 and No.~11575017;
the Chinese Academy of Science Center for Excellence in Particle Physics; 
the Ministry of Education, Youth and Sports of the Czech
Republic under Contract No.~LG14034;
the Carl Zeiss Foundation, the Deutsche Forschungsgemeinschaft, the
Excellence Cluster Universe, and the VolkswagenStiftung;
the Department of Science and Technology of India; 
the Istituto Nazionale di Fisica Nucleare of Italy; 
the WCU program of the Ministry of Education, National Research Foundation (NRF) 
of Korea Grants No.~2011-0029457,  No.~2012-0008143,  
No.~2014R1A2A2A01005286, 
No.~2014R1A2A2A01002734, No.~2015R1A2A2A01003280,
No.~2015H1A2A1033649, No.~2016R1D1A1B01010135, No.~2016K1A3A7A09005603, No.~2016K1A3A7A09005604, No.~2016R1D1A1B02012900,
No.~2016K1A3A7A09005606, No.~NRF-2013K1A3A7A06056592; 
the Brain Korea 21-Plus program and Radiation Science Research Institute;
the Polish Ministry of Science and Higher Education and 
the National Science Center;
the Ministry of Education and Science of the Russian Federation and
the Russian Foundation for Basic Research;
the Slovenian Research Agency;
Ikerbasque, Basque Foundation for Science and
the Euskal Herriko Unibertsitatea (UPV/EHU) under program UFI 11/55 (Spain);
the Swiss National Science Foundation; 
the Ministry of Education and the Ministry of Science and Technology of Taiwan;
and the U.S.\ Department of Energy and the National Science Foundation.


\begin{thebibliography}{99}
\bibitem{CC}
  Throughout this paper, the inclusion of the charge conjugate mode
  decay is implied unless stated otherwise.
\bibitem{natural-units}
  The use of natural units ($c=1$) is assumed throughout the paper.
\bibitem{combined}
  For the combined sample, $\mathcal{B}(\eta)$ is the sum of the $\eta\to\gamma\gamma$ and $\eta\to\pi^+\pi^-\pi^0$
  branching fractions.
\end{thebibliography}
\end{document}